# Characterization of irradiated microstructure by X-ray diffraction line profile analysis

A. SARKAR, P. MUKHERJEE and P. BARAT*

Variable Energy Cyclotron Centre

1/AF Bidhan Nagar, Kolkata 700064, India

*Corresponding author, email: pbarat@veccal.ernet.in

**ABSTRACT**

Zirconium based alloys have been irradiated with 11 and 15 MeV proton and 116 MeV oxygen ions at different doses. The changes in the microstructure have been studied for the ion irradiated alloys as a function of dose using X-Ray Diffraction Line Profile Analysis (XRDLPA) based on the whole powder pattern fitting technique. It was observed that the microstructural parameters like domain size, microstrain within the domain, dislocation density did not change significantly with the increase in dose for proton irradiated samples. A clear change was noticed in these microstructural parameters as a function of dose for oxygen irradiated samples. There was a drastic decrease in domain size at a dose of $1\times10^{17}$ $O^{5+}/m^2$ but these values reached a plateau with increasing dose. The values of microstrain and dislocation density increased significantly with the dose of irradiation.

## I. INTRODUCTION

Changes in the properties of materials by irradiation are of great technological interest. The structural materials used in the core of the nuclear reactors experience significant changes during operation due to the change in the microstructure, phase distributions, dimensions, electrical properties, magnetic properties and corrosion [1, 2] under severe radiation environment. Keeping the technological interests in view, several studies have been carried out on the problems of simulating fission and fusion neutron damage with light and heavy ions [3-7]. Extensive studies have also been carried out on the nuclear structural materials [8, 9] on the response of the light and heavy particle irradiation and consequent defect production and microstructural evolution.

In the early stages of the projectile trajectory, the collision sequences are widely spaced [10], giving a distribution of single vacancies and interstitials. Towards the end of the projectile trajectory, the distance between the collisions decreases to the inter-atomic distance. This generates a large number of Primary Knock-on Atoms (PKA) causing a highly localized production of vacancies and interstitials. Hence the distribution of vacancies and interstitials within the damaged regions is non-uniform [11]. The presence of substitutional and interstitial impurities can conceivably result in a change in the total amount of damage at a given dose by two mechanisms, firstly, replacement collision sequences may be terminated and the defect production rate will decrease by the enhancement of mutual recombination and secondly the trapping of point defects may result in the retention of point defect clusters which might otherwise be lost by mutual recombination on migration to sinks [2, 12]. These result in an overall change of the microstructure forming vacancy clusters, dislocation loops or small domains, immobile clusters of self interstitials etc [2, 13]. The irradiation induced defect clusters play a major role in governing the mechanical properties of the structural materials. The structural materials used inside the nuclear reactors are in cold-worked and heat treated conditions. The damage accumulation and the microstructural evolution during irradiation affect the in-reactor behavior particularly irradiation creep and irradiation growth which finally determine the structural integrity of these materials during service. Still today in-situ characterization of irradiated microstructural parameters is in its infancy and post-irradiation examination of microstructure and evaluation of other mechanical properties are carried out inside hot-cell due to severity of radiation environment. Moreover, for the predictions of material properties under reactor conditions, we need long irradiation times to be able to reliably predict neutron-induced changes. Hence, comes the need of ion-irradiation. By using light or heavy ions and varying their incident energies, the recoil spectrum can be altered so that it covers the significant ranges of neutron recoil spectrum [1]. The principal environmental parameters include the type and energy of the incident particle, the fluence which is connected to the total number of displacements produced in the material, the displacement rate i.e. dpa/sec, the pre-irradiation microstructure and composition of the material [1].

We have characterized the microstructure of Zircaloy-2 and Zr-1Nb-1Sn-0.1Fe alloy which have been irradiated with proton at 11 MeV, 15 MeV and 116 MeV $O^{5+}$ ions. The main problem associated with ion-irradiation is the inhomogeneous damage profile, as a result, the irradiated microstructure will be spatially heterogeneous. Hence, the study of statistically averaged microstructural parameters and their variation with increasing dose of irradiation are of utmost importance to understand the mechanism of radiation damage. X-Ray Diffraction Line Profile Analysis (XRDLPA) evaluates the microstructural parameters in a statistical manner averaged over a volume of $10^9$ $\mu m^3$. Hence, the bulk damage can be assessed by this analysis. With the advent of the computer based programs for the microstructural characterization by profile fitting, XRDLPA has become an immensely powerful technique [14].

The basic mechanism behind the diffraction of X-ray in crystalline materials is that X-rays get scattered from crystals since their electric fields interact with the electron clouds of the atoms in the crystals. The scattered X-rays from the periodic adjacent atoms interfere and give rise to diffraction pattern. The diffraction pattern is modulated by the transfer function of the detector which in turn changes the shape of the X-ray diffraction profile. Thus, a diffraction line profile is the result of the convolution of a number of independent variables contributing to shape profile viz instrumental variables and microstructural effects. Instrumental variables include receiving slit width, sample transparency, the nature of the X-ray source, axial divergence of the incident beam and flat specimen geometry [15]. The microstructural effects that are responsible for the shape profile of the diffraction peaks are the finite size of the crystals or domains and the microstrain within the domain as the crystal contain lattice defects. These profiles are fitted with suitable profile shape functions in such a way that the functions must fit the asymmetric peaks and it should be mathematically as simple as possible to make the calculation of all derivatives to the variables. Normally the Pearson VII function, Voigt function and pseudo-Voigt (pV) function are used to fit the experimental curves [16].

We have characterized the microstructure of the irradiated zirconium based alloys using different techniques of XRDLPA [17, 18]. Williamson-Hall technique, Double Voigt technique, Integral Breadth method on single and multiple peaks and Modified Rietveld technique have been used to characterize the microstructural parameters like domain size, microstrain within the domain, dislocation density of the irradiated material as a function of dose. In our earlier studies [14, 17-20], the mathematical formalism of the different techniques have been discussed in detail. Here, we have analyzed the diffraction peaks based on the recently developed whole powder pattern fitting technique using the program MarqX [21]. The microstructural parameters have been assessed from the Size-Strain plot (SS plot) [22] and the Warren-Averbach (WA) analysis [23, 24].

The prime objective of this paper is to compare the effect of light and heavy ion irradiation on the microstructure of zirconium based alloys and to understand the variation of the microstructural parameters caused by these ions in the light of the mechanism of defect evolution and their distribution in the matrix.

## II. EXPERIMENTAL PROCEDURE

Ingots of Zircaloy-2 and Zr-1Nb-1Sn-0.1Fe alloy were prepared in the Nuclear Fuel Complex, Hyderabad, India, by the double vacuum arc melting technique. It was then β quenched, followed by hot extrusion and cold pilgering to produce fuel cladding tubes of 0.4 mm wall thickness.

Samples of size 10 mm × 10 mm were cut from these tubes of both the alloys. Samples of Zr-1Nb-1Sn-0.1Fe alloy were annealed at 1023 K for 4 h. The samples were mounted on an aluminum flange. The samples of Zircaloy-2 were irradiated with 11 MeV proton and the annealed samples of Zr-1Nb-1Sn-0.1Fe were irradiated by 15 MeV proton and 116 MeV $O^{5+}$ ions from Variable Energy Cyclotron (VEC), Kolkata, India. The irradiation doses were $9.85\times10^{21}$, $5.0\times10^{22}$ protons/$m^2$ for 11 MeV proton, $1\times10^{20}$, $5\times10^{20}$, $5\times10^{21}$ protons/$m^2$ for 15 MeV proton and $1x10^{17}$, $5x10^{17}$, $1x10^{18}$ and $5x10^{18}$ $O^{5+}$ ions/$m^2$ for 116 MeV $O^{5+}$ ion. The flange used for irradiation was cooled by continuous flow of water. During the irradiation, the temperature of the sample did not rise above 313K as measured by the thermocouple connected very close to the sample. The range of ions in this material and the damage profiles were obtained by Monte-Carlo simulation technique using the code SRIM 2000 [25].

X-Ray Diffraction (XRD) profiles for each irradiated sample have been recorded from PHILIPS 1710 diffractometer using CuK$_{\alpha}$ radiation. The range of 2θ was from 30° to 95° and a step scan of 0.02° was used. The time per step was 4 seconds.

## III. METHOD OF ANALYSIS

MarqX [21] is a computer program for the modeling of diffraction patterns; it mostly addresses problems of materials science which typically include lattice-parameter determination and line profile analysis.

The relation between the instrumental profile component (g), the profile broadening due to domain size and lattice defects (f) and the experimentally observed profile (h) is [26]

$$h(x)=f(x-y)\otimes g(y) \qquad (1)$$

In this computer program it is assumed that h, f, g can be properly modeled by pV function. The g profile can be obtained by means of a suitable standard material ($LaB_6$ for the present analysis), from which the instrumental resolution function (IRF) can be modeled (22, 27). In particular, MarqX employs the following two expressions for parametric description of the IRF:

$$(2HWHM)^2 = U\tan^2\theta + V\tan\theta + W \qquad (2a)$$

$$\eta = a\theta + b \qquad (2b)$$

where $U$, $V$, $W$, $a$, $b$ are refinable parameters, HWHM is half width at half maximum, $\theta$ is the Bragg angle and $\eta$ is the mixing parameter of the pV function. Since the analytical profiles are pV functions, the integral breadth ($\beta$) can be obtained as

$$\beta_{pV} = HWHM[(1-\eta)(\pi/\ln 2)^{1/2} + \eta\pi] \qquad (3)$$

The knowledge of HWHM, $\eta$, $\beta$ permits the calculation of Voigt parameter $\phi = \frac{2HWHM}{\beta}$ [22]. The corresponding Lorentz and Gauss components of the integral breadth ($\beta_{gL}$, $\beta_{fL}$, $\beta_{gG}$, $\beta_{fG}$ respectively) can be calculated from [28]

$$\beta_{hL} = \beta_{gL} + \beta_{fL} \qquad (4a)$$

$$\beta_{hG} = (\beta_{gG}^2 + \beta_{fG}^2)^{1/2} \qquad (4b)$$

where $\beta_{hL}$, $\beta_{hG}$ are the Lorentz and Gauss component of the total $\beta$. $\beta_{gL}$, $\beta_{fl}$ are the Lorentz components of instrument and sample contribution respectively and $\beta_{gG}$, $\beta_{fG}$ are the corresponding Gauss components.

Once the parameters in equation (2a and 2b) are known from the analysis of a suitable profile of a standard material, the convolutive approach permits a direct refinement of the pV parameters of the f-profile, which can then be used in the line profile analysis.

Using this program, the Williamson-Hall plot (WH plot) i.e. $\beta^* = \frac{\beta\cos\theta}{\lambda}$ vs $d^* = \frac{2d\sin\theta}{\lambda}$, where $d$ is the interplanar spacing, $\lambda$ is the wavelength of the X-ray, and the SS plot i.e. $\left(\frac{\beta^*}{d^*}\right)^2 vs \frac{\beta^*}{(d^*)^2}$ [22] can be obtained. Moreover, profile parameters are used to calculate the Fourier coefficient ($A_L$) of each modeled peak corrected for instrumental broadening. The WA plot i.e. $\ln(A_L)$ vs. $(d^*)^2$ is plotted to separate size and strain component according to WA method [23,24,29]. WA analysis can be performed on all the reflections assuming isotropy or by selecting specific (hkl) reflection.

Considering the X-ray line profiles to be symmetric in shape, distributions of the dislocations were assumed to be random. The average density of dislocation (ρ) has been estimated from the relation [30] $\rho = (\rho_D \rho_S)^{\frac{1}{2}}$, where, $\rho_D = \frac{3}{D_s^2}$ (density of dislocation due to domain) and $\rho_S = k\langle \varepsilon_L^2 \rangle / b^2$ (density of dislocation due to strain), k is the material constant, $b$ is the modulus of the Burger's vector, $\frac{1}{3}[11\bar{2}0]$, $D_s$ is the surface weighted domain size and $\varepsilon_L$ is the microstrain within the domain.

**IV. RESULTS AND DISCUSSIONS**

Fig 1 shows the damage profiles of 11 MeV proton in Zircaloy-2, 15 MeV proton and 116 MeV $O^{5+}$ beam in Zr-1Sn-1Nb-0.1Fe obtained using the program SRIM 2000. The range of 11 MeV and 15 MeV proton in Zircaloy-2 and Zr-1Sn-1Nb-0.1Fe was found to be approximately 440 μm and 742 μm respectively. In both the cases the range of the projectile is more than the thickness (400 μm) of the samples irradiated. Hence, the proton completely penetrated the sample and only a fraction of the elastic energy of the proton beam was deposited on it. Thus, the damage produced in the samples was a bulk phenomenon due to complete penetration of the proton beam. On the contrary, the range of 116 MeV $O^{5+}$ beam in Zr-1Sn-1Nb-0.1Fe is around 67 μm as calculated by SRIM 2000. Oxygen being a heavy ion impart so much energy to the primary knock on atoms that a displacement cascade is produced consisting of highly localized production of interstitials and vacancies associated with a single initiated event. In contrast, the reaction pathways of proton beam through the sample of thickness 400 μm produces fairly uniform radiation damage. Moreover, the damage energy deposition with distance i.e. dE/dx is found to be almost two orders less than that of oxygen beam as shown in Fig. 2 (calculated from the code SRIM 2000).Thus, during the travel of oxygen ion the total elastic energy deposited within the material of 67 μm is much larger than that of proton. Besides, as the primary recoil proceeds through the sample, loosing energy in successive collisions, the displacement cross-section increases [31]. Thus the distance between successive displacements decreases [31] and at the end of its track, the recoil collides with practically every atom in its path, creating a very high local concentration of vacancies and interstitials.

In our earlier observations [32] on Zr–1Nb–1Sn–0.1Fe samples irradiated with proton of 15 MeV, we have seen by positron annihilation spectroscopy studies that there was a generation of irradiation induced di-vacancy and tri-vacancy clusters. Hence, the domain size, microstrain values and the order of dislocation density remained almost unchanged even up to the highest dose of irradiation [17], as seen by XRDLPA using modified Rietveld technique. We further concluded that proton irradiation at these doses on Zr–1Nb–1Sn–0.1Fe samples did not generate additional dislocation loops through collapse of point defects.

Fig 3 shows typical XRD profiles of unirradiated and irradiated (dose $5.0\times10^{22}$ protons/$m^2$, irradiated with 11 MeV proton) Zircaloy-2 samples. Fig. 4 represents typical XRD profiles for Zr–1Nb–1Sn–0.1Fe samples irradiated with 116 MeV $O^{5+}$ beam (dose $5.0\times10^{18}$ $O^{5+}/m^2$) along with the unirradiated one. Fig 3 clearly depicts that the broadening is almost negligible for proton irradiated sample as compared to the unirradiated one but clear broadening is observed for oxygen irradiated sample as shown in Fig. 4. Fig. 5 represents a typical whole powder pattern fit using MarqX for the XRD profile of oxygen irradiated Zr–1Nb–1Sn–0.1Fe sample at a dose of $5.0\times10^{17}$ $O^{5+}/m^2$. Typical figures of SS plot, WA plot, $A_s(L)$ vs. L plot and microstrain vs. L plot (L being the Fourier length) are shown in Fig. 6 and Fig. 7 for proton irradiated Zircaloy-2 sample (dose $5.0\times10^{22}$ protons/$m^2$, irradiated with 11 MeV proton) and oxygen irradiated Zr–1Nb–1Sn–0.1Fe sample (dose $5.0\times10^{17}$ $O^{5+}/m^2$) respectively. The volume weighted domain size ($D_v$) and microstrain are obtained from the SS plot. The surface weighted domain size ($D_s$) is obtained from the initial slope of the $A_s(L)$ vs. L plot.

Table 1a, 1b, 2a, 2b and 3a, 3b give the results obtained by fitting the profiles at different doses using the whole powder pattern fitting technique followed by SS plots and WA analysis respectively on the proton irradiated and oxygen irradiated samples. We observed that the $D_s$, $D_v$ and microstrain did not change significantly even with increasing dose of irradiation for proton irradiated samples which corroborated with our earlier results on Zr–1Nb–1Sn–0.1Fe irradiated by 15 MeV proton, analyzed by modified Rietveld technique [17]. The values of dislocation density also remained unaltered with the increase in the dose of irradiation as observed in Fig. 8a and 8b. On the contrary, we find significant changes in the domain size and microstrain with dose in oxygen-irradiated samples as compared to the unirradiated one as shown in Table 3a and 3b. There is a significant decrease in domain size with dose as compared to the un-irradiated sample but these values saturate with increasing dose of irradiation. The values of microstrain are found to increase with dose. The dislocation density increases almost by an order of magnitude for the samples irradiated with $5\times10^{17}$, $1\times10^{18}$ and $5\times10^{18}$ $O^{5+}/m^2$

as compared to the unirradiated sample as shown in Fig. 8c. These values are also found to saturate with dose. The reasons of the above findings can be explained as follows:

As seen from the damage profile of 11 MeV and 15 MeV proton (Fig. 1a and 1b), the Bragg peak of the target displacements falls beyond the thickness of the samples which means that only a fraction of the total elastic energy is being utilized for causing displacement of atoms and as a result isolated point defects are created in the irradiated samples of zirconium based alloys. These isolated point defects do not cause broadening of the peaks or change in the shape of the profile but contribute to the background values close to the diffraction peaks due to diffuse scattering (Huang scattering). As a result the domain size, microstrain and dislocation density as observed in the unirradiated sample remain almost unaltered even with increasing dose of irradiation in case of proton irradiated samples.

For $O^{5+}$ ion irradiation, the damage is maximum within a distance of 2-3 μm at the end of the reaction path. A concentration gradient of defects in the sample was thus created within a small reaction path of 67μm which helped in the migration of defects. Moreover, the diffusion coefficient $D_a$ of a particular lattice atom is enhanced due to irradiation [33] and is given by the following equation:

$$D_a = f_v D_v C_v + f_{2v} D_{2v} C_{2v} + f_i D_i C_i + ...... \quad (5)$$

where *v*, *2v* and *i* stand for vacancy, di-vacancy and interstitial respectively. *D* and *C* values are the corresponding diffusion coefficients and concentrations. *f* values are the correlation factors [33].

The migration of defects by radiation enhanced diffusion resulted agglomeration into defect clusters which collapsed in the form of dislocation loops. With increasing dose of irradiation, though more vacancies are created, the annihilation rate of vacancies also increases with increasing sink density [33]. Hence, a saturation was observed in the density of dislocation with the increase in the dose of irradiation.

## V. CONCLUSION

In conclusion, we have established that XRDLPA based on the whole powder pattern fitting technique can be reliably used to study the variation of the microstructure caused by light and heavy ion irradiation in Zirconium based alloys. The effects of light and heavy ions on the defect evolution and their distributions in the matrix have been distinguished distinctly by this technique.

**REFRENCES**

Table 1a Results obtained from Size-Strain (SS) plot of the XRD profiles of Zircaloy-2 irradiated by 11 MeV proton.

| Dose (protons/$m^2$) | $D_v$ (Å) | Microstrain ($10^{-3}$) |
|---|---|---|
| Unirradiated | 231±12 | 1.2±0.3 |
| $9.85\times10^{21}$ | 249±22 | 1.3±0.4 |
| $5.0\times10^{22}$ | 237±10 | 1.1±0.2 |

Table 1b Results of Warren-Averbach (WA) analysis of the XRD profiles of Zircaloy-2 irradiated by 11 MeV proton.

| Dose (protons/$m^2$) | $D_s$ (Å) | Microstrain ($10^{-3}$) |
|---|---|---|
| Unirradiated | 181±7 | 1.3±0.4 |
| $9.85\times10^{21}$ | 189±11 | 1.4±0.6 |
| $5.0\times10^{22}$ | 172±5 | 1.2±0.3 |

Table 2a Results obtained from Size-Strain (SS) plot of the XRD profiles of Zr–1Nb–1Sn–0.1Fe irradiated by 15 MeV proton.

| Dose (protons/$m^2$) | $D_v$ (Å) | Microstrain ($10^{-4}$) |
|---|---|---|
| Unirradiated | 1547±63 | 7.23 |
| $1\times10^{20}$ | 1449±86 | 7.66 |
| $5\times10^{20}$ | 1532±76 | 7.02 |
| $5\times10^{21}$ | 1517±79 | 7.99 |

Table 2b Results of Warren-Averbach (WA) analysis of the XRD profiles of Zr–1Nb–1Sn–0.1Fe irradiated by 15 MeV proton.

| Dose(protons/$m^2$) | $D_s$ (Å) | Microstrain ($10^{-4}$) |
|---|---|---|
| Unirradiated | 971±52 | 8.6 |
| $1\times10^{20}$ | 923±49 | 6.9 |
| $5\times10^{20}$ | 986±56 | 7.3 |
| $5\times10^{21}$ | 1012±53 | 7.7 |

Table 3a Results obtained from Size-Strain (SS) plot of the XRD profiles of Zr–1Nb–1Sn–0.1Fe irradiated by 116 MeV $O^{5+}$ ion.

| Dose ($O^{5+}/m^2$) | $D_v$ (Å) | Microstrain ($10^{-3}$) |
|---|---|---|
| Unirradiated | 1547±63 | 0.72 |
| $1x10^{17}$ | 603±51 | 1.12 |
| $5x10^{17}$ | 442±44 | 1.44 |
| $1x10^{18}$ | 411±46 | 1.79 |
| $5x10^{18}$ | 387±35 | 1.96 |

Table 3b Results of Warren-Averbach (WA) analysis of the XRD profiles of Zr–1Nb–1Sn–0.1Fe irradiated by 116 MeV $O^{5+}$ ion.

| Dose ($O^{5+}/m^2$) | $D_s$ (Å) | Microstrain ($10^{-3}$) |
|---|---|---|
| Unirradiated | 971±52 | 0.86 |
| $1x10^{17}$ | 431±29 | 1.01 |
| $5x10^{17}$ | 337±31 | 1.55 |
| $1x10^{18}$ | 288±23 | 1.82 |
| $5x10^{18}$ | 268±19 | 2.03 |

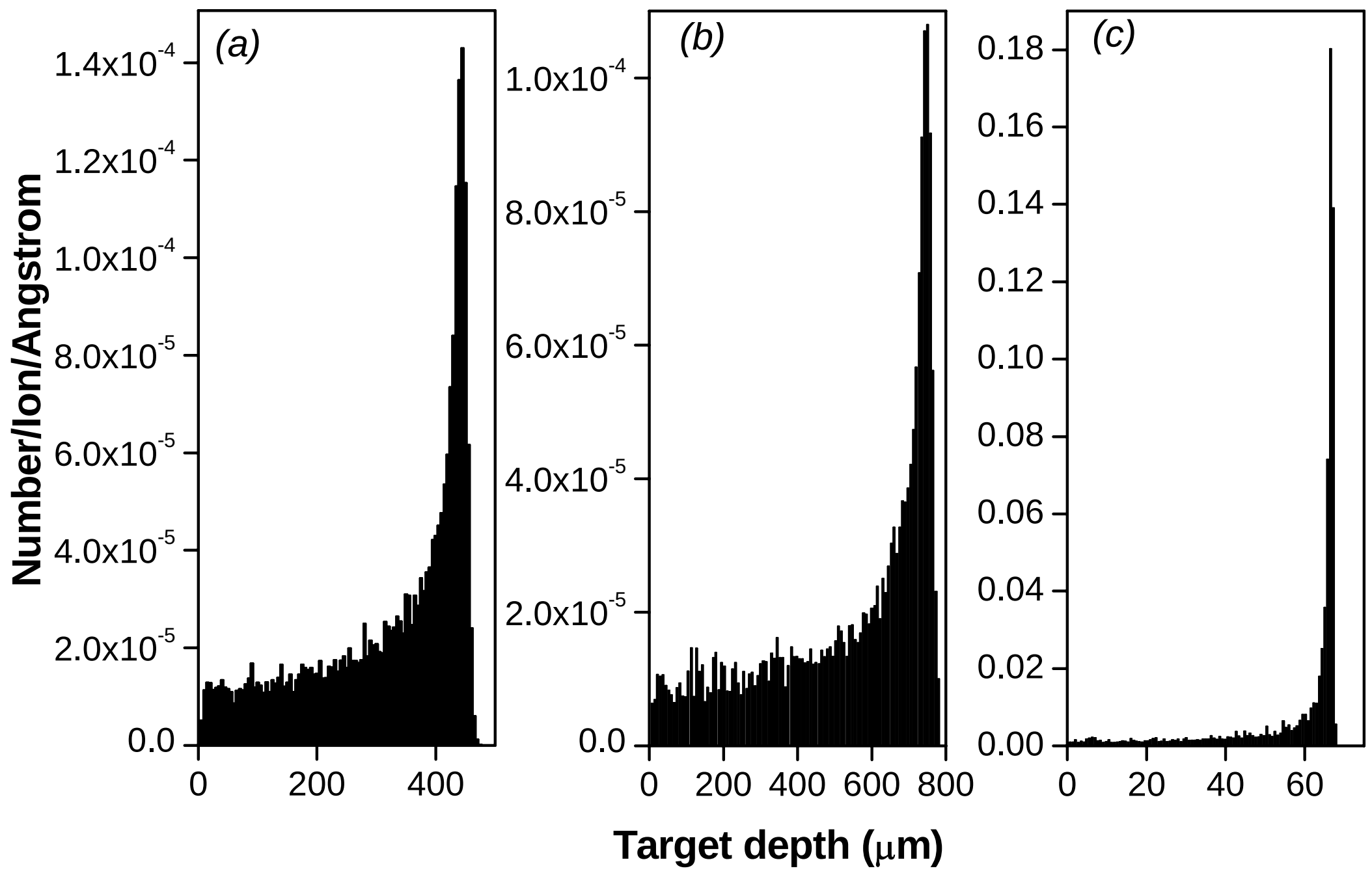

Fig. 1 Damage profile of (a) 11MeV proton in Zircaloy-2 (b) 15 MeV proton in Zr-1Nb-1Sn-0.1Fe (c) 116 MeV $O^{5+}$ in Zr-1Nb-1Sn-Fe.

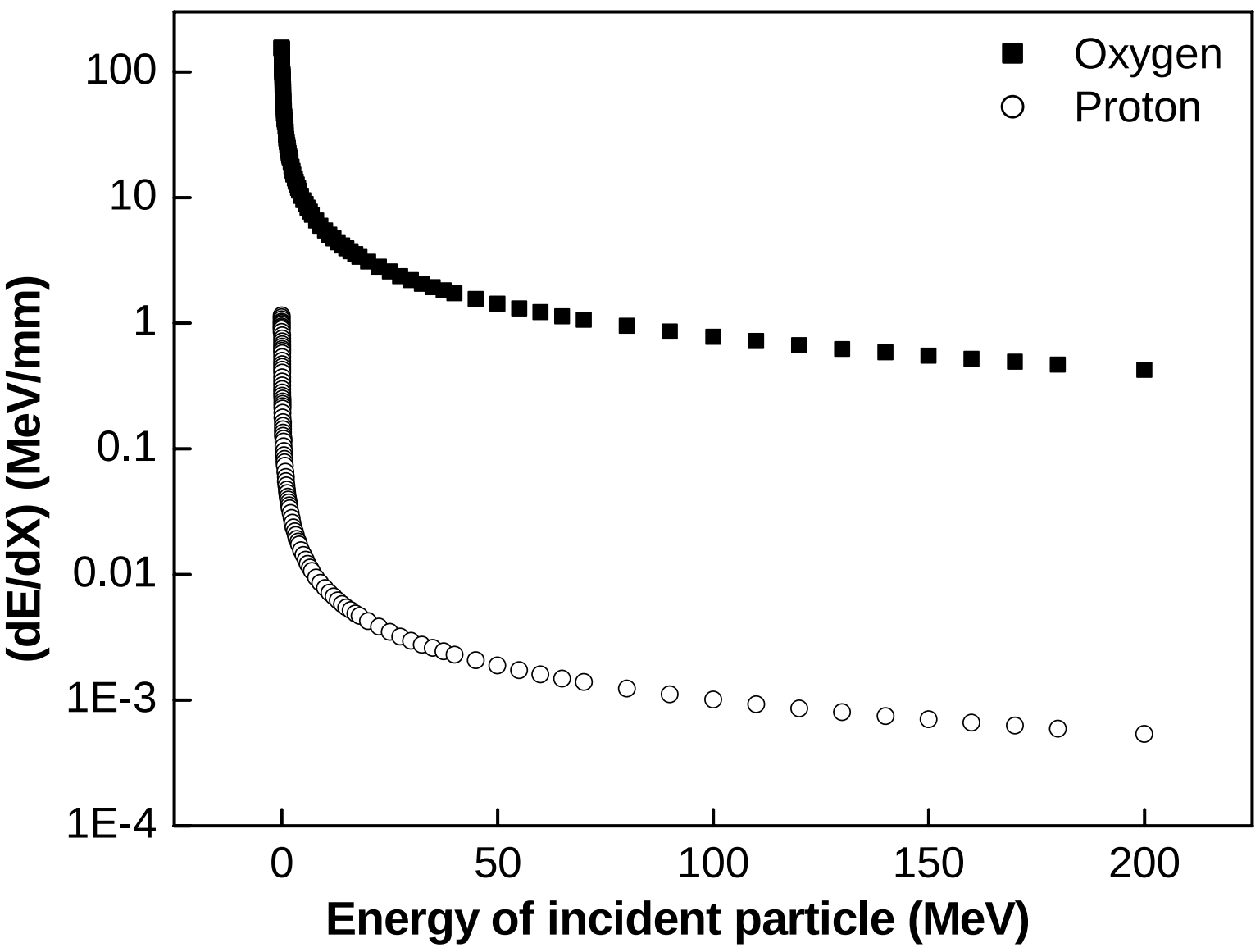

Fig. 2 Variation of (dE/dx) with respect to energy of the incident particle (MeV)

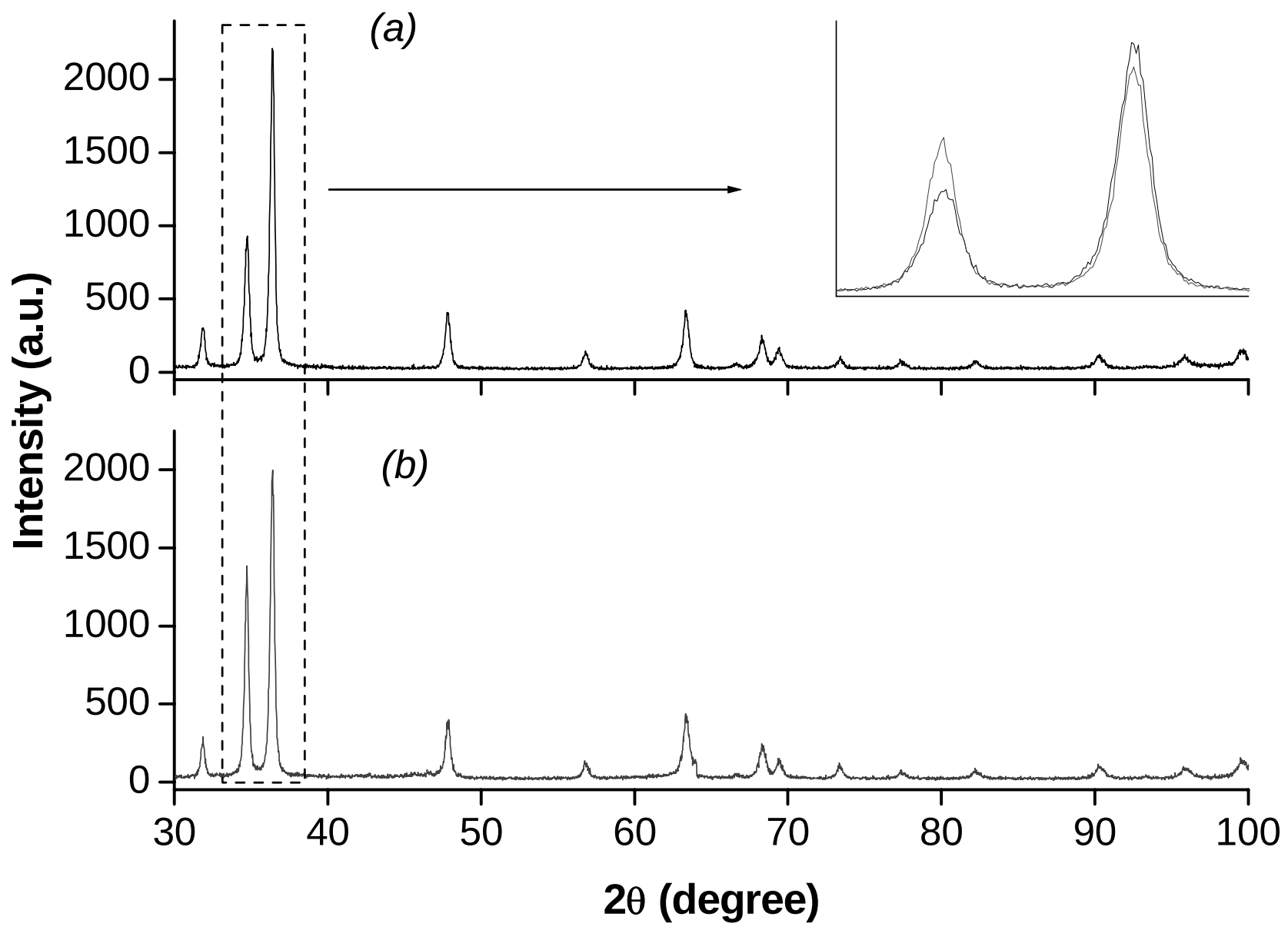

Fig. 3 XRD profile of the Zircaloy-2 (a) unirradiated (b) irradiated at dose $5.0\times10^{22}$ protons/$m^2$. Inset shows the segments of the two XRD profile.

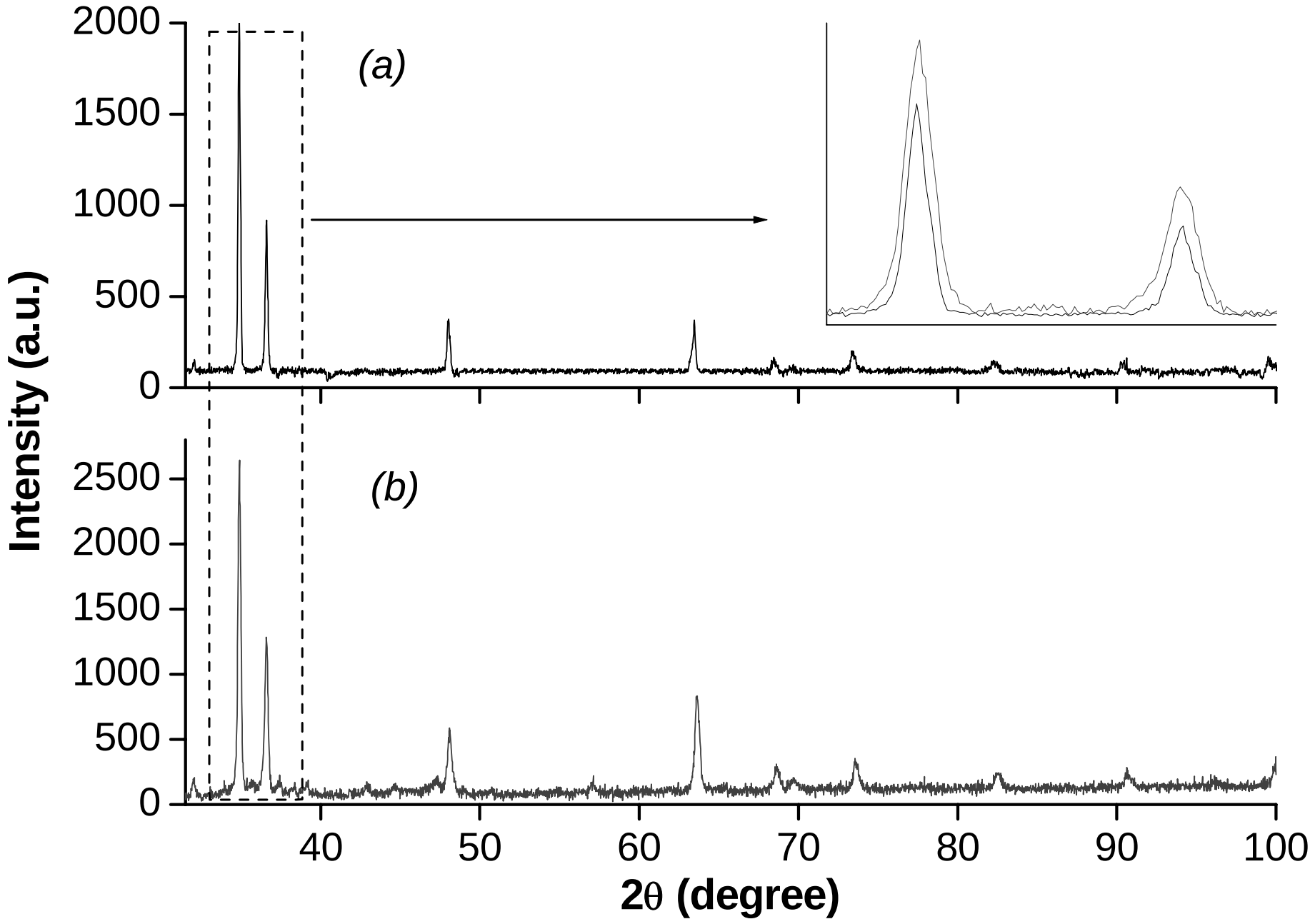

Fig. 4 XRD profile of the Zr-1Nb-1Sn-0.1Fe (a) unirradiated (b) irradiated at dose $5.0\times10^{18}$ $O^{5+}/m^2$. Inset shows the segments of the two XRD profile.

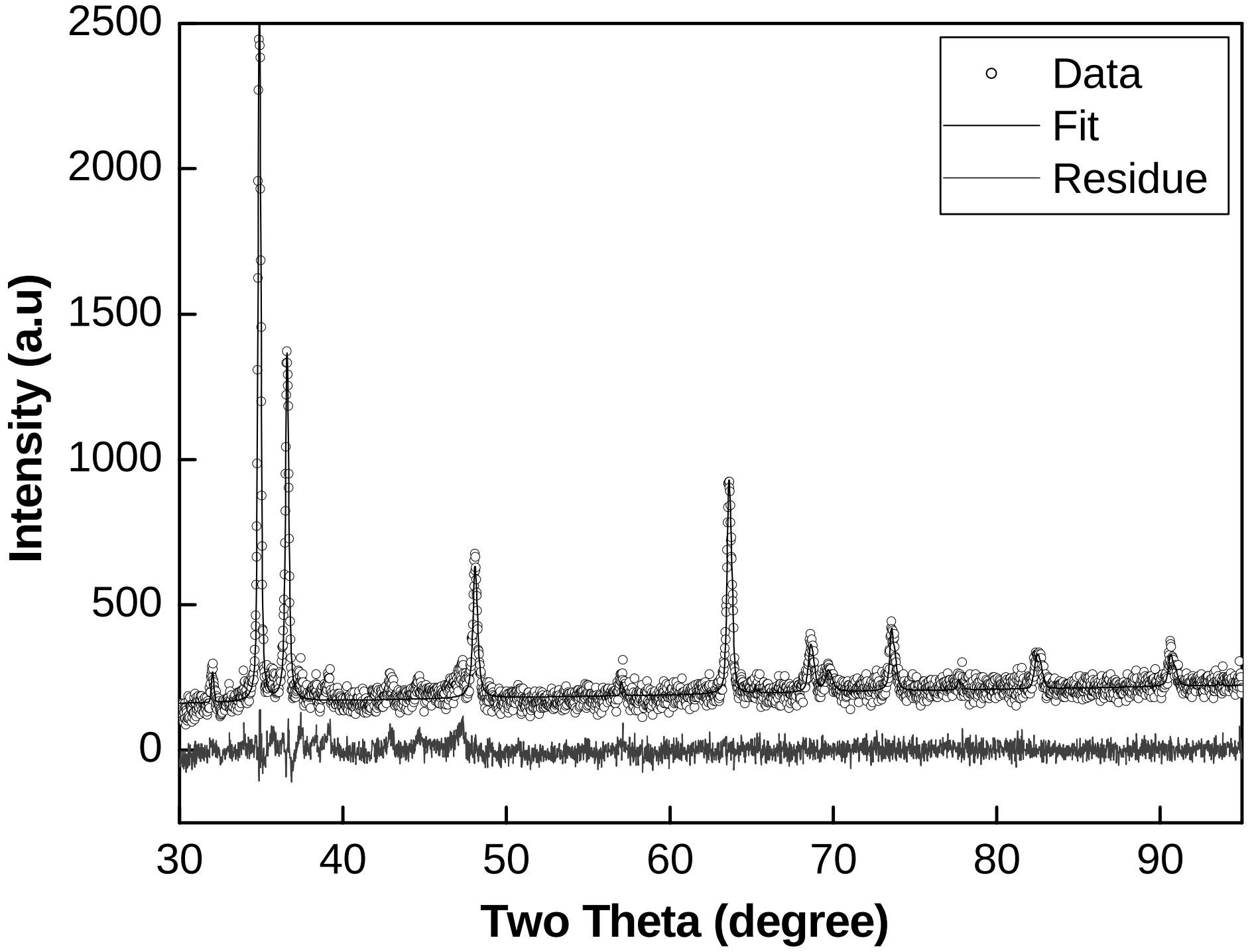

Fig. 5 Whole powder pattern fit for oxygen irradiated Zr-1Nb-1Sn-0.1Fe at dose $5.0\times10^{17}$ $O^{5+}/m^2$.

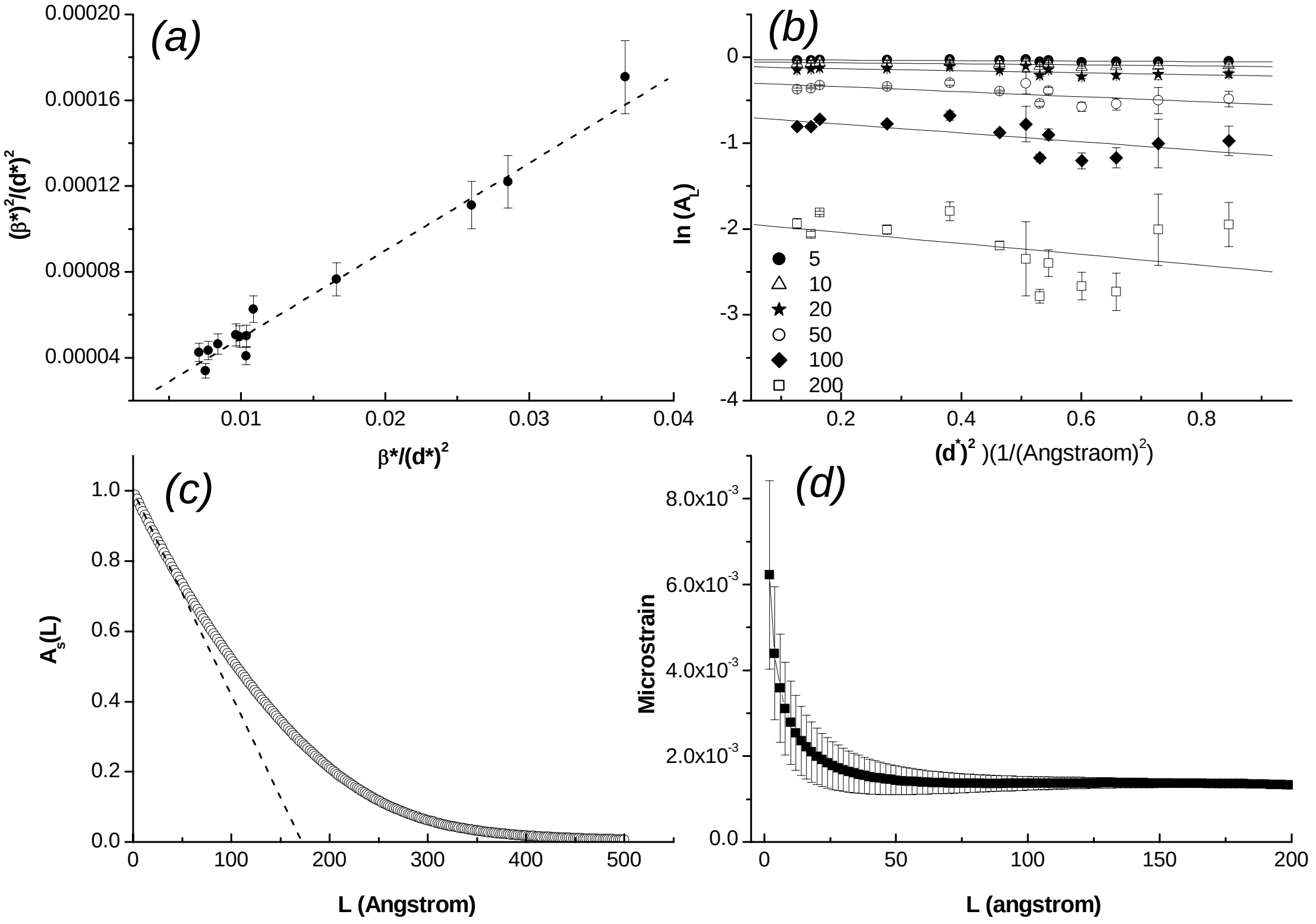

Fig. 6 (a) SS plot (b) WA plot (c) $A_s(L)$ vs L plot (d) microstrain vs L plot for proton irradiated (dose $5.0\times10^{22}$proton/m$^2$) Zircaloy-2.

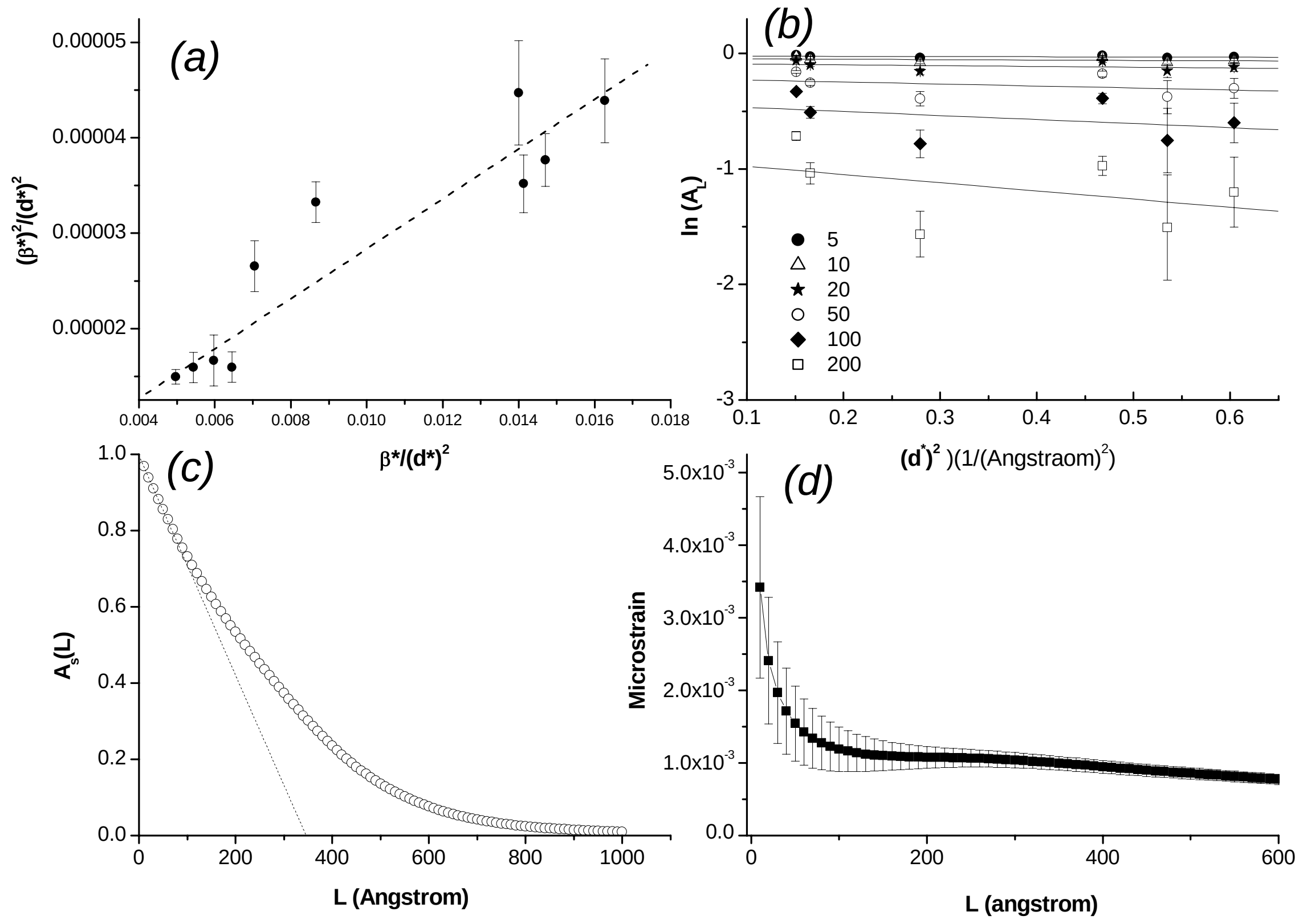

Fig. 7 (a) SS plot (b) WA plot (c) $A_s$(L) vs L plot (d) microstrain vs L plot for oxygen irradiated (dose $5.0\times10^{17}$ $O^{5+}/m^2$) Zr-1Nb-1Sn-0.1Fe.

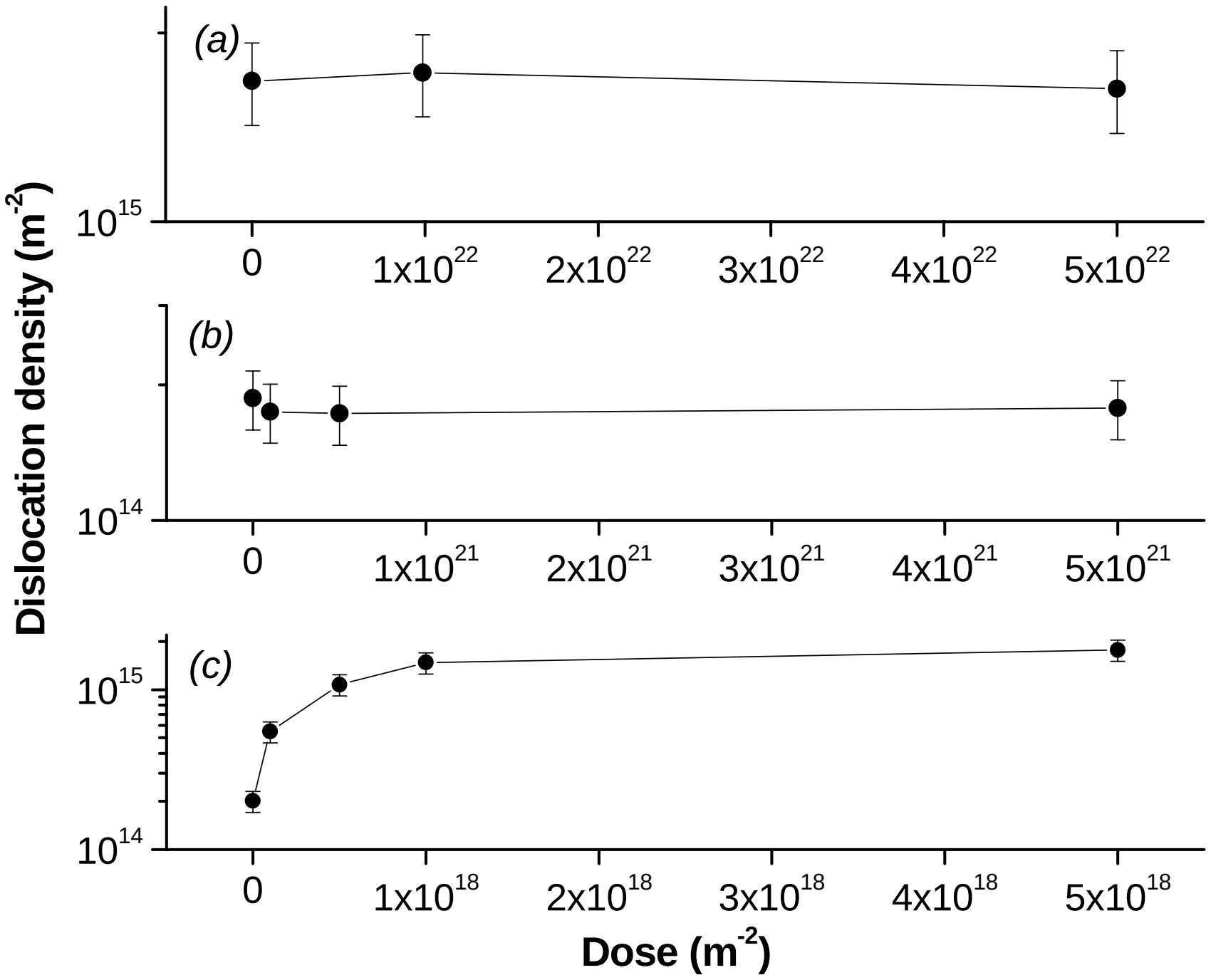

Fig. 8 Variation of dislocation density with dose of irradiation for (a) Zircaloy-2 irradiated with 11 MeV proton, (b) Zr-1Nb-1Sn-0.1Fe irradiated with 15 MeV proton (c) Zr-1Nb-1Sn-0.1Fe irradiated with 116 MeV $O^{5+}$ ion.